\documentclass[prb,twocolumn,showpacs,eqsecnum]{revtex4}

\usepackage{graphicx,graphics,color,epsfig}% Include figure files
\usepackage{bm}
\usepackage{amsmath}
\usepackage{amssymb}
\usepackage{epstopdf}
\usepackage{euscript}
\usepackage{longtable}

\newcommand{\beq}{\begin{equation}}
\newcommand{\eeq}{\end{equation}}
\newcommand{\beqn}{\begin{eqnarray}}
\newcommand{\eeqn}{\end{eqnarray}}
\begin{document}
%\draft

%\preprint{submitted to Phys. Rev. B}
\title{ Coulomb correlation in presence of spin-orbit coupling: application to plutonium}
\author{Jean-Pierre Julien}
\affiliation{Theoretical Division, Los Alamos National Laboratory, MS B262,  Los Alamos, NM 87545, USA}
\affiliation{Institut Neel CNRS and Universit\'{e} J. Fourier 25 Avenue des Martyrs, BP 166, F-38042
Grenoble Cedex 9, France}
\author{Jian-Xin Zhu}
\affiliation{Theoretical Division, Los Alamos National Laboratory, MS B262,  Los Alamos, NM 87545, USA }
\author{R. C. Albers}
\affiliation{Theoretical Division, Los Alamos National Laboratory, MS B262,  Los Alamos, NM 87545, USA }

\date{\today}

\begin{abstract}
Attempts to go beyond the local density approximation  (LDA) of
Density Functional Theory (DFT) have been increasingly based on the
incorporation of more realistic Coulomb interactions.
In their earliest implementations,
methods like LDA+$U$, LDA + DMFT
(Dynamical Mean Field Theory), and LDA+Gutzwiller used
a simple model interaction $U$. In this
article we generalize the solution of the full Coulomb matrix
involving $F^{(0)}$ to $F^{(6)}$ parameters, which is usually
presented in terms of an $\ell m_\ell$ basis, into a  $jm_{j}$ basis
of the total angular momentum, where we also
include spin-orbit coupling;
this type of theory is needed for a
reliable description of $f$-state elements like plutonium,
which we use as an example of our theory.
Close attention will be paid to spin-flip terms, which are important in multiplet theory
but that have been usually neglected in these kinds of studies.
We find that, in a density-density approximation,
the $jm_j$ basis results provide a very good approximation to the full Coulomb
matrix result, in contrast to the much less accurate results for
the more conventional $\ell m_\ell$ basis.
\end{abstract}
\maketitle

\section{\label{s:int}Introduction}

Strongly correlated electron systems are solids
where the important outer-shell electrons have two conflicting and
opposite tendencies. On one hand, they maintain a strong memory
of the atomic or localized orbitals from which they arise, which have a large
electron-electron electrostatic interaction between discrete states.
On the other hand, the same electrons hybridize with neighboring orbitals
causing them to delocalize by
tunnelling from one atom to those nearby, forming
chemical bonds and spreading out the discrete atomic states
into narrow energy bands.
In such systems, it is therefore necessary for
a correct description and
understanding of their electronic properties to maintain
both of these aspects.
The second tendency is very accurately calculated by density
functional theory (DFT) band-structure calculations, while the first
one involves a consideration of many-body effects, and, while more
difficult to treat, is still a crucial aspect of the physics.
Thus, it is important to increase our knowledge
of the Coulomb interactions that strongly affect the atomic
character of these systems, an effect which is often underestimated or poorly
approximated in calculations that include details of the band-structure.

These types of effects are particularly important for the
electronic structure of $f$-electron elements in general, and especially for the
actinides. For these materials, density functional theory (DFT)
calculations in the local density
approximation (LDA)~\cite{Hohenberg64,Kohn65} often give significant discrepancies.
For example,
$\delta$-Pu from this kind of approach is predicted to have an equilibrium volume 25\% smaller
than experiment, which is the largest known deviation from LDA.
To overcome these difficulties, various attempts to go
beyond LDA have been proposed, such as LDA+$U$,~\cite{Anisimov91,
Bouchet00}
LDA+DMFT,~ \cite{Georges96,Anisimov97,Savrasov00} and
more recently LDA+Gutzwiller.~\cite{JulienBouchet06} All three of these
methods add a local Hubbard-like term to a band Hamiltonian,
and require subtraction of an average LDA Coulomb
interaction (the double counting correction).
The differences between the various methods reside in the way the
effects of this interaction term is handled.

In the LDA+$U$ approach, which employs a Hartree-Fock mean-field
solution, the Hubbard term
leads to an orbital-dependent shift in the potential.
Such a crude mean-field approximation is questionable
for cases involving strong correlations.
In the LDA+Gutzwiller method,
a variational wavefunction is built,
for which the mean values of the interaction are calculated exactly.
In the more sophisticated DMFT method, the effect of the interaction is
described by a self-energy, which acts as an energy-dependent
complex potential. In this approximation the self-energy is assumed to be
local (i.e., momentum independent) and is determined
self-consistently within an impurity-like approach of a
correlated site embedded in a effective bath.  In many DMFT calculations
the Hubbard-like term, at least in the early implementation of these
methods, has often been treated in a fairly simplistic way.
For example, some applications use
a single $U$ term, average over all interactions, while others also include
an exchange-averaged parameter $J$. Over time the general tendency for all
three methods has been to include more and more realistic interactions.
For example, since its first use, the LDA+$U$ has usually been
rotationally invariant.~\cite{rotinvLDA+U} However, a multiband version of
Gutzwiller approach developped by B\"{u}nemann, Gebhard and
Weber~\cite{Bunemann98b} was able to handle spin-flip terms and very
recently, a multiband generalization of slave-boson
formalism~\cite{Kotliar86} has been proposed to be rotationally
invariant.~\cite{rotinvSB}
To make progress, it is clearly important to develop a
more sophisticated treatment of the Coulomb interactions.
In addition, for high-Z materials like the $f$-electron actinides,
spin-orbit must also be accurately included.
To do this is the goal of this article.

The paper is organized as follows. Section II is devoted to the
presentation of Coulomb matrix elements in the $j m_j$ basis. In
this section, we first formulate a general expression for the
interactions. We then show how to make an approximate
density-density correlation calculation for these interactions. The
corresponding matrix elements will be tabulated in terms of
Slater integrals. Section III presents the eigen-spectrum
of the atomic Hamiltonian in various different approximations.
In Section IV we use these eigenvalues to study the single particle
spectral density. In particular, the quality of various approximations
will be evaluated against a rigorous solution.
Finally, we conclude in Section V by stressing the main
results of our approach.

\section{\label{s:coul}Coulomb matrix elements in the $j m_j$ basis}

\subsection{Electronic structure of an isolated atom}

The Hamiltonian $H$ of an isolated many-electron atom or ion is
\beq H=\sum_{i=1}^{N} \biggl{(}\frac{-\hbar^{2}}{2m}\nabla_{i}^{2}-\frac{Z
e^{2}}{r_i}+\xi(r_{i})
\overrightarrow{L_{i}}\cdot \overrightarrow{S_{i}}\biggr{)}+
\sum_{i>j}\frac{e^{2}}{|r_i-r_j|}\label{Hexact}\;. \eeq
Beside the interaction of the electrons with the positive
charge (Z) of the nucleus, this Hamiltonian contains
two important features: the
spin-orbit coupling and the electrostatic (Coulomb) interaction
between electrons. By neglecting spin-orbit, in the central field
approximation,~\cite{CondonShortley}
the eigenstates of
the system are Slater determinants built from individual states
$|n \ell m_\ell s m_s \rangle$ having the following wave
function:
 \beq
\phi_{n \ell m_{\ell} s m_{s}}(r,\theta, \varphi)= \frac{R_{n
\ell}(r)}{r} Y_{\ell m_{\ell}}(\theta, \varphi)\eta_{sm_{s}}(s_z)\;. \eeq
Here $Y_{\ell m_{\ell}}$ is a spherical harmonics,  $R_{n \ell}$ the
solution of a radial Schr\"{o}dinger equation, and $\eta_{sm_{s}}(s_{z})$
an eigenfunction of $S_{z}$.
%Together with
%individual spin $ s=\frac{1}{2}$ and its projection
%$m_{s}=\pm\frac{1}{2}$,
The set of indices $ n \ell  m_{\ell} s
m_{s}$ is sufficient to determine completely a state with
eigenvalue $E_{n \ell}$, having the degeneracy $2(2\ell+1)$. It is
this basis (or its equivalent in a $jm_j$ basis)
that we will use to study the full Hamiltonian
(\ref{Hexact}). The appearance of a two-body term, i.e., the
electrostatic interaction between electrons,  makes the
problem sufficiently complicated so that an eigenstate, even in a
perturbative description, will not be in general be a single Slater
determinant.

The spin-orbit coupling, which is still a one-body operator, is a relativistic
effect and can be directly obtained in the Schr\"{o}dinger
formulation as a limit of the Dirac equation. It is due to the
interaction of the magnetic moment of electron spin with
the effective magnetic field created by the orbital motion, and has the
following expression:
\beq H_{so}= \sum_{i=1}^{N} \xi(r_{i})
\overrightarrow{L_{i}}\overrightarrow{S_{i}}\;,
 \eeq
with,  by dropping the index $i$,
\beq  \xi(r)=\frac{1}{2m^{2}c^{2}r}\frac{\partial V}{\partial r}\;.
\label{xi} \eeq
The spin-orbit interaction is diagonal in a $j m_j$ basis and splits the
$j=\ell \pm \frac{1}{2}$ states into two subsets having correction
energies  $\Delta_{\ell + \frac{1}{2}} =\ell \chi_{n
\ell}/2$ and $\Delta_{\ell - \frac{1}{2}}= -(\ell+1) \chi_{n
\ell}/2 $, respectively.  The splitting energy is  $(2\ell+1) \chi_{n \ell}/2$. Here
$\chi_{n \ell}$ is given by the radial integral:
\beq
\chi_{n \ell}= \hbar^{2} \int  \xi(r) |R_{n\ell}(r)|^2  dr \;.
\label{SOsplitting}
\eeq
Spin-orbit coupling begins to be important for atoms with
atomic number $Z\geq 20$ , where the derivative $\frac{\partial
V}{\partial r}$ starts to become significant.

\subsection{Two limiting behaviors: $LS$ or $jj$ couplings}

Depending on which of two contributions, the Coulomb
or spin-orbit interaction, dominates, there are two
limiting regimes. The first is the $LS$ coupling or
Russell-Saunders regime, in which the electrostatic  exchange
interaction is predominant; this is responsible for the Hund's rule
ordering of states. In this case $\ell m_\ell$ is the most convenient
basis with the unperturbed states in the
form $|n \ell m_\ell s m_s \rangle$, since the Coulomb matrices are
diagonal in spin.

The opposite regime, the $jj$ coupling regime, occurs when
spin-orbit coupling splitting is greater than the electrostatic
terms. In  that case, it is convenient to work in the $j m_j$
basis, which diagonalizes the spin-orbit term and to treat the
electrostatic term as a first-order pertubation, leading to a
single (diagonal) correction to the unperturbed eigenenergies.

For actinides we are in an intermediate regime. For example, the
average exchange $J$ in plutonium \cite{Shick06} is of the order
of 0.7eV and the spin-orbit parameter $\chi_{5f}$ for $5f$ states
is in the range 0.25-0.54 eV, producing an energy splitting between
$j=5/2$ and $j=7/2$ states in the range of 0.9-1.95 eV. In this case,
it is important to treat the spin-orbit
and the electrostatic terms on the same footing by diagonalizing
them in a given
basis. Transformation from one basis to the other  can be
performed with the use of the Clebsch-Gordan coefficients
$\langle \ell \, m_\ell \, s \, m_s|j \, m_j \rangle$ with
\beq |j \, m_j\rangle=\sum |\ell \, m_\ell  \, s \, m_s \rangle
\langle \ell \, m_\ell \, s \, m_s | j \, m_j\rangle \;. \eeq
As we will explain below, there are strong arguments for using the
the $j m_j$ basis.  In this case,
the diagonal part gives directly the $jj$ coupling
approximation and the fully diagonalized result provides a
reliable description of the intermediate regime.

\subsection{Coulomb interaction in the $jm_j$ basis}

In the $j m_j$ basis we can write the Coulomb contribution to the Hamiltonian as
\beq V_{C}= \frac{1}{2}
\sum_{1234}V_{1234}c_{1}^{\dagger}c_{2}^{\dagger}c_{4}c_{3}
\label{VC} \eeq
Here 1, 2, 3, and 4 are a shorthand for individual particle
states $|n  \ell j m_j \rangle$.
% It can be an eigenstate of the angular
%momentum $L^2$ and $L_z$ and spin $S^2$ and $S_z$ operators in the
%decoupled atomic basis $|\ell \; m_\ell,s \; m_s \rangle \bigotimes
%|\frac{1}{2} \sigma\rangle \equiv |l_i m_l,\frac{1}{2} \sigma>$. This
%state has also a radial part, such that the wavefunction
%$\langle \overrightarrow{r} \sigma|i>=\Psi_{i}(\overrightarrow{r})$. It
%can be also an eigenstate $|j_i m_i>$ of $J^2$ and $J_z$ in the
%coupled basis
%($\overrightarrow{J}=\overrightarrow{L}+\overrightarrow{S}$ is the
%total momentum).
The spatial part of the Coulomb interaction can be expanded
in the $\ell m_\ell$ basis as
\beq
\frac{1}{|r-r'|}\\
=\sum_{k=0}^{\infty}\sum_{q=-k}^{k}\frac{r_{<}^k}{r_{>}^{k+1}}\frac{4\pi}{2k+1}
Y_{k,q}^{*}(\Omega)Y_{k,q}(\Omega')~. \label{coulexpans}\;\eeq
Here $r_{<}$
($r_{>}$) is the lesser (greater) of $r$
and $r'$, and $Y_{k,q}$ is a spherical harmonics, with
 the solid angle spanned by $\Omega=(\theta,\varphi)$.

The Coulomb matrix element $V_{1234}$ is explicitly given by
(using the system of units where $e^{2}=1$)
\beq V_{1234}=\langle j_1
m_1 j_2 m_2| \frac{1}{|r-r'|}| j_3 m_3 j_4 m_4\rangle
\label{V1234}\;.\eeq
In this expression different
levels of approximation can be made. If it is used exactly as is with no
approximation, we will refer to the results as involving
``spin-flip'' terms,
since the creation and destruction operator in $V_{C}$
can flip spins. We will call the
next level of approximation the ``density-density correlation" approximation
since, as it will become clear below,
in this approximation one retains
only the case for which either $1\equiv3$ and $2\equiv4$ (the
direct term) or $1\equiv4$ and $2\equiv3$ (the exchange term).
Thus this part of the Hamiltonian reduces to \beq V_{C}= \frac{1}{2}
\sum_{1234}(V_{1212}-V_{1221})n_{1}n_{2} \eeq It is worth noting
that the usual selection rule that occurs in the  $\ell m_\ell$ basis,
namely, that the exchange interaction vanishes for antiparallel spins,
does not occur here, since the $j m_j$ basis
is a mixture of different $\ell
m_{\ell}$ and $s m_s$ states.  As a result, in our present case, a
net interaction within the density-density correlation
approximation is always the difference between a direct and an
exchange term.

The density-density correlation approximation is very important
because it make the LDA+$U$ feasible. For the case of DMFT, if one uses the
Hirsch-Fye type~\cite{Hirsch-Fye86} Quantum Monte Carlo (QMC) solver,
the Hubbard-Stratonovitch transformation makes this
approximation necessary.  For the Gutzwiller case, even if it is in
principle possible to keep the spin-flip terms,~\cite{Bunemann98b}
it is however much easier to avoid these terms and to make the
density-density correlation approximation, especially for the
density-matrix derivation of the generalized Gutzwiller
method.~\cite{JulienBouchet06}

$ $

\subsubsection{\label{ss:gener}General formulation including spin-flip terms}

Our starting point is the definition (\ref{V1234}).  Since we are mainly
interested in $f$-electron elements ($\ell=3$), the possible  values of the
$j=\ell\pm\frac{1}{2}$ in this expression is either 7/2 or 5/2. In
the Coulomb potential expansion (\ref{coulexpans}), it is suitable
to insert a closure relation in the decoupled basis, and make
further use of selection rules of Clebsch-Gordan coefficients. As
a result, we obtain (see Appendix for a detailed demonstration):
%\widetext
%\begin{eqnarray}
\begin{align}
\langle j_1 m_1  j_2 m_2&|V| j_3 m_3 j_4 m_4\rangle=
(-)^{m_3-m_2} \delta_{m_1+m_2,m_3+m_4}       \nonumber\\
&\times \sum_{k=0}^{2\ell} F^{(k)}\frac{(2\ell+1)^2}{(2k+1)^2}
 \langle \ell \, 0 \, \ell \, 0 | k \, 0\rangle^{2}     \nonumber\\
&\times B_{k}^{\ell}(j_1 m_1;j_3 m_3) %\nonumber \\
B_{k}^{\ell}(j_2 m_2; j_4 m_4)\;,
\label{Vspflip}
\end{align}
%\end{eqnarray}
%\endwidetext
\noindent where the summation over $k$ extends only over even values of $k$
ranging from $0$ to $2l$ due to selection rule for the
Clebsch-Gordan coefficient $\langle \, \ell \, 0 \, \ell \, 0| k \, 0\rangle$ (see Appendix).
The $B_{k}^{\ell}(jm;j' m')$ are given by
%\widetext
%\begin{eqnarray}
\begin{align}
B_{k}^{\ell}&(jm;j' m')= \sum_{\sigma=\pm\frac{1}{2}} 2\sigma
\langle \ell \: (m{-}\sigma) \: \frac{1}{2} \: \sigma | jm\rangle \nonumber\\
&\times
\langle \ell \: (m'{-}\sigma) \: \frac{1}{2} \: \sigma | j' m' \rangle \nonumber\\
&\times
\langle \ell \:  {-}(m{-}\sigma) \: \ell \: (m'{-}\sigma) | k \:
(m'{-}m)\rangle \label{B}\;,
\end{align}
%\end{eqnarray}
%\endwidetext
and the $F^k$ are the Slater integrals
 \beq
 F^{(k)}=\int  dr dr'|R_{n\ell}(r)|^2 \frac{r_{<}^k}{r_{>}^{k+1}}|R_{n\ell}(r')|^2~.
\label{SlaterInt}
 \eeq

\subsubsection{\label{ss:dens-dens}
Density-density correlation approximation}

As explained above, in the density-density correlation we retain
among all possible terms of (\ref{Vspflip}) only those for which
$(j_1 m_1) \equiv (j_3 m_3)$ and $(j_2 m_2)\equiv (j_4 m_4)$ for the direct
term, whereas $(j_1 m_1)\equiv  (j_4 m_4)$ and $(j_2 m_2)\equiv (j_3 m_3)$
gives the exchange term. By using (\ref{Vspflip}) with these
restrictions, we obtain these two kinds of matrix elements:
\begin{eqnarray}
U_{m_1 m_2}^{j_1 j_2}&=& \langle j_1 m_1 \; j_2 m_2| V|j_1 m_1 \; j_2 m_2\rangle\;, \nonumber\\
J_{m_1 m_2}^{j_1 j_2}&=& \langle j_1 m_1 \; j_2 m_2| V|j_2 m_2 \;
j_1 m_1\rangle \;,
\label{Vddapp}\end{eqnarray}
which are found to be
\begin{eqnarray}
U_{m_1 m_2}^{j_1 j_2}&=& \sum_{k=0}^{2\ell} a_{k}^{\ell}(j_1 m_1;
j_2 m_2) F^{(k)} \;,\nonumber\\
J_{m_1 m_2}^{j_1 j_2}&=& \sum_{k=0}^{2\ell} b_{k}^{\ell}(j_1 m_1;
j_2 m_2)F^{(k)} \;,
\label{Vddapp2}
\end{eqnarray}
where
%\widetext
\begin{eqnarray}
a_{k}^{\ell}(j_1 m_1; j_2 m_2) =
&&(-)^{m_2-m_1}\frac{(2\ell+1)^2}{(2k+1)^2} \langle \ell \, 0 \, \ell \,
0 | k \, 0\rangle^{2}\nonumber\\
&&\times B_{k}^{\ell}(j_1 m_1;j_1 m_1) B_{k}^{\ell}(j_2 m_2;j_2 m_2)\;,
\nonumber\\
\label{aso}
\end{eqnarray}
%\endwidetext
and
%\widetext
\begin{eqnarray}
b_{k}^{\ell}(j_1 m_1; j_2 m_2)= &&\frac{(2\ell+1)^2}{(2k+1)^2}
\langle \ell \, 0 \,  \ell \, 0 | k \, 0\rangle^{2}\nonumber\\
&&\times B_{k}^{\ell}(j_1 m_1; j_2 m_2)^{2}\;. \label{bso}
\end{eqnarray}
%\endwidetext

We have checked that we can use these two formulas for
$\ell\leq 2$ to retrieve results first established by
Inglis \cite{Inglis31} in the early 1930's and which are now common
in textbooks (see, for example, Ref.~\onlinecite{CondonShortley}).
Since results for $f$-electron elements ($\ell=3$) in these references are not
provided,  we have tabulated them in Tables \ref{tab:direct} and
\ref{tab:exchange} of the Appendix, where
we give also some symmetry relations they obey. These results are
a generalization for the $j m_j$ basis of the more familiar $\ell
m_{\ell}$ results:
\begin{eqnarray}
U_{m m'}^{\ell \ell}&=& \sum_{k=0}^{2\ell} a_k F^{(k)}\;,\nonumber\\
J_{m m'}^{\ell \ell}&=& \sum_{k=0}^{2\ell} b_k F^{(k)}\;,
\end{eqnarray}
with
\begin{eqnarray}
a_k&=& \frac{4\pi}{2k+1} \sum_{q=-k}^{+k} \langle Y_{\ell m}|Y_{k
q}^*|Y_{\ell m}\rangle \langle Y_{\ell m'}|Y_{k q}|Y_{\ell
m'}\rangle \;,\nonumber\\
b_k&=& \frac{4\pi}{2k+1} \sum_{q=-k}^{+k} |\langle Y_{\ell m}|Y_{k
q}|Y_{\ell m'}\rangle|^2 \;.
\end{eqnarray}

\subsubsection{\label{ss:shibrec} Averaged interactions and the single $U$ limit}

The direct calculation of $F^{(k)}$ from the
atomic wave functions in their definition (\ref{SlaterInt})
usually overestimates their values, since it neglects screening
effects that occur in the solid, \textit{i.e.}, the relaxation of
other electrons pushed away by the two interacting electrons,
which leaves behind a net positive charge and reduces the strength of the
interaction. Consequently, effective values for $F^{(k)}$ are
usually computed
from constrained LSDA (local spin-density approximation) DFT
calculations.\cite{AnisimovGunnarsson91} Unfortunately, because LSDA can
not distinguish between individual orbitals, it is only
sensitive to the total spin-density and hence this approach can only provide an
averaged direct and exchange interactions.

For example, if this average is performed for the $\ell m_\ell$
basis, one obtains for $f$-states the following averaged direct and
exchange interactions\cite{SawatzkyCzyzyk}
 \begin{eqnarray}
 \overline{U}&=&\frac{1}{(2\ell+1)^2}\sum_{m m'} U_{m m'} \nonumber \\
 &=& F^{(0)}\;,
 \label{avUlm}
 \end{eqnarray}
and
 \begin{eqnarray}
\overline{U}-\overline{J}&=&\frac{1}{2\ell(2\ell+1)}\sum_{m m'}
(U_{m m'}-J_{m m'})\nonumber\\
&=& F^{(0)}-(286F^{(2)}+195F^{(4)}+250F^{(6)})/6435\;. \nonumber \\
\label{avJlm}
\end{eqnarray}
>From the constrained LSDA values of $\overline{U}$ $(=F^{(0)})$ and
$\overline{J}$, and requiring constant ratios
of $F^{(6)}/F^{(2)}$ and $F^{(4)}/F^{(2)}$,
 it is possible to assign a unique value to each $F^{(k)}$ (k=0, 2, 4, 6).
This procedure, when applied to plutonium, leads to the values given in
Table \ref{tab:slater}.~\cite{Shick06}

We have generalized the same kind of average for the
direct and exchange interaction in the $j m_j$ basis for
$f$ states, and find:
 \beq \overline{U}=\frac{1}{(2j_1+1)(2j_2+1)}\sum_{m m'} U_{m m'}^{j_1 j_2}
 =F^{(0)}\label{avUjmj}\;,\eeq
and
 \beq
 \overline{J}_{j_1 j_2}=\frac{1}{(2j_1+1-\delta_{j_1 j_2})(2j_2+1)}\sum_{m m'} J_{m m'}^{j_1 j_2}\;.
%U-J=\frac{1}{2\ell(2\ell+1)}\sum_{m m'} (U_{m m'}-J_{m
%m'})=F^{(0)}-(F^{(2)}+F^{(4)
\label{avJjmj}\eeq
The exchange terms vary if they are taken for
different $j_{1}{-}j_{2}$ pairs (5/2-5/2, 5/2-7/2, and
7/2-7/2). Accordingly, one finds:
\begin{eqnarray}
\overline{J}_{5/2 \; 5/2}&=& \frac{1}{30}(\frac{48}{35}F^{(2)} + \frac{4}{7}F^{(4)})\;,\nonumber\\
\overline{J}_{5/2 \; 7/2}&=& \overline{J}_{7/2 \;
5/2}=\frac{1}{48}(\frac{8}{25}F^{(2)} +
 \frac{40}{77}F^{(4)}+\frac{200}{143}F^{(6)})\;,
\nonumber\\
\overline{J}_{7/2 \; 7/2}&=&
\frac{1}{56}(\frac{40}{21}F^{(2)}+\frac{72}{77}F^{(4)}+\frac{200}{429}F^{(6)})\;.
\end{eqnarray}
If we go one step further in the averaging process and use a
single $\overline{J}$ value, regardless of which $j$ (5/2 or 7/2)
it comes from, then we obtain:
\begin{eqnarray}
\overline{J}=\frac{1}{182}(\frac{56}{32}F^{(2)}+\frac{28}{11}F^{(4)}+\frac{1400}{429}F^{(6)})\;.
\label{Jbar}
\end{eqnarray}

These values can be used to further approximate the
density-density approach, since the obvious simplification is to
replace the detailed interaction given by (\ref{Vddapp}) by a
single value, which is their average over all possible pairs of orbitals.
This last approximation has been widely used and we discuss
next its effect on the spectrum.

%\begin{center}
%Table I  \\
\begin{table}
%\vspace{.5cm}
\caption{\label{tab:slater}Slater integral
values (in eV) for Plutonium from Ref.~\onlinecite{Shick06}.}
\begin{ruledtabular}
\begin{tabular}{cccc}
$F^{(0)}$ & $F^{(2)}$ & $F^{(4)}$ & $F^{(6)}$ \\
\hline
4.   &  8.343639   &  5.57482 & 4.12446 \\
\end{tabular}
\end{ruledtabular}
\end{table}
%\end{center}
%\vspace{1cm}
%\begin{center}

\section{\label{s:ecm} Spectrum of eigenvalues}

\subsection{Hamiltonian and Fock states}

We now explain how to obtain the eigenvalues of the atomic
local part of the Hamiltonian (i.e., with no kinetic energy term)
in second quantization form:
\begin{eqnarray} H_{loc}&=& \frac{1}{2}\sum_{1234} V_{1234}
c_{1}^{\dagger}c_{2}^{\dagger}c_{4}c_{3} + \sum_{1}
\Delta_{1}c_{1}^{\dagger}c_{1}\;,
\end{eqnarray}
where $p$ (=1, 2, 3, or 4) is a shorthand for $j_p m_p$,
$V_{1234}$ is the Coulomb potential (see Eq.~\ref{V1234}), and $\Delta_{1}$
is the spin-orbit term (see just below Eq.~\ref{xi}),
which is diagonal in the $jm_j$ basis.

First of all, this Hamiltonian conserves the number of particles.
Thus, for a given occupancy $N$ ($0 \leq N \leq 14$) of
$f$-levels, the dimension of the basis for configuration $f^N$ is
${14!}/{N!(14-N)!}$, which is the number of Fock states
$|n_N\rangle$ for spreading N electrons among the $i=1$ to $14$
individual quantum states arising from the six 5/2 states (with
$m_j$ ranging from -5/2 to +5/2) and the eight 7/2 states (with $m_j$
ranging from -7/2 to +7/2):
\begin{equation} \vert n_N\rangle=
(c_{1}^{\dagger})^{o_1}...(c_{14}^{\dagger})^{o_{14}}\vert 0
\rangle \;.\label{Fockstate} \end{equation}

Each state of the form (\ref{Fockstate}) in this subspace has an
electron occupancy of $N$,
\textit{i.e.}, $\sum o_i=N $, where $o_i$ is either 0 or 1
in order to obey the Pauli principle. We compute then the matrix elements
$\langle n'_N|H|n_N \rangle$ of the Hamiltonian in this Fock basis of
states. The diagonalization of this matrix gives the energies
$E^{(N)}_{a}$ of the multiplet structure and their corresponding
eigenstates $|a\rangle$, with the Fock state components:
\begin{equation}
 \vert a \rangle=\sum_{n_N} \langle n_N  \vert a\rangle  \vert n_N
\rangle\;.
\end{equation}
 As mentioned above, if only the diagonal part of Hamiltonian,
which corresponds to density-density approximation, is
kept (valid in the limit
of the $jj$ coupling), the Fock states are already eigenstates of the system,
and the eigenvalues are the diagonal matrix elements.

Moreover, as we are dealing with isolated atoms, the eigenvalues
are discrete. Thus, to plot the spectrum, we have added to each eigenvalue
a small imaginary part, which modifies the spectra from a sum of
$\delta$ functions to a sum of equal-width Lorentzians (here
set equal to $0.1\;\text{eV}$).

%\newpage
\begin{figure}[h]
%\begin{center}
%\rotatebox{0}{\includegraphics[width=8cm]{fig1.eps}} \vspace{.5cm}
\centerline{\psfig{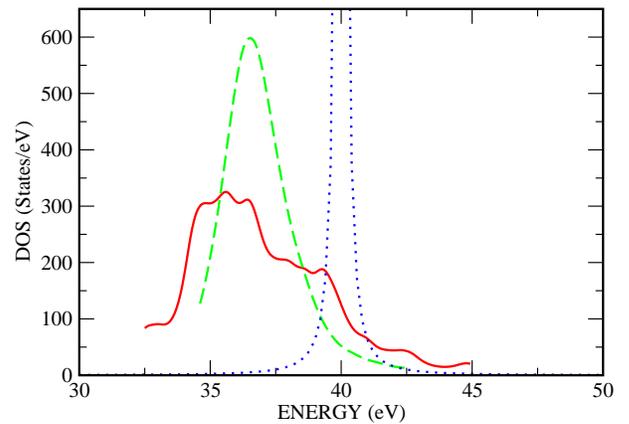}}
\caption{\label{spectr-noLS} (Color online) Atomic spectrum of
eigenvalues of Pu without spin-orbit coupling, with N=5 electrons
for single U (dotted line), density-density approximation (dashed
line) and spin-flip terms (full line) included respectively. }
%\end{center}
\end{figure}
%\newpage

\subsection{Results}
\begin{figure}[t!]
%\begin{center}
%\rotatebox{0}{\includegraphics[width=8cm]{fig2.eps}} \vspace{.5cm}
\centerline{\psfig{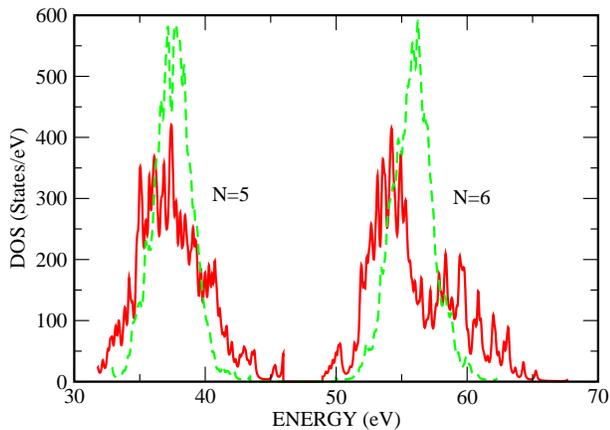}}
\caption{\label{spectr+LS} (Color online) Atomic spectrum of
eigenvalues of Pu with spin-orbit coupling, with N=5 and N=6
electrons for density-density approximation (dashed line) and
spin-flip terms included (full line) respectively. }
%\end{center}
\end{figure}

In Figs.~\ref{spectr-noLS} and \ref{spectr+LS}, the density of
states (DOS) is the sum of all Lorentzians from each eigenvalue.
We note that this DOS \textit{should not} be confused with the
spectral density presented in next part: it is just a visual way
of presenting the distribution of eigenvalues. For illustration,
we present the case of plutonium where the Slater integrals
parameters are given in Table I. Figure~\ref{spectr-noLS} presents
the atomic spectrum of eigenvalues of Pu without spin-orbit
coupling, for a given occupancy of $N=5$ electrons. The different
approximations with increasing order of complexity are displayed:
single $U$, density-density approximation and spin-flip terms
included respectively. Figure \ref{spectr+LS} displays the same
situation when spin-orbit coupling has been taken into account
for occupancies of $N=5$  and $N=6$ electrons.

The most accurate description is given when the spin-flip terms
are not neglected. In this case, the spectrum is much more
structured than the result for the density-density approximation.
The spin-flip terms in the Hamiltonian couple different states
that are degenerate in the
density-density approximation, and split them.
This causes the additional
structure in the spectrum. The density-density
approximation spectrum therefore has less
structure, but has its gross features centered in the
same region as the spectrum which include the spin-flips.
For the single $U$ limit result, for a
given occupancy $N$ and neglecting of spin-orbit coupling, there is a
single eigenvalue $U {N(N-1)}/{2}$ whose degeneracy equals the
dimension of subspace. This is the origin, when smeared with a Lorentzian,
of the single peak, which is actually too high in
energy and well separated from the more realistic spectrum in the
density-density approximation or the one including spin-flip terms
(roughly 5 eV above the center of gravity of the other ones).
Clearly, the single $U$ limit overestimates the interaction
energy. This is mainly caused by the complete neglect of any
exchange term, which would reduce the average interaction. This
approximation, which has been widely used, can be a problem for an
atom embedded in a solid, since the metal-insulator
transition is a delicate balance between localization due
to the interaction $U$ and delocalization, represented by a bandwidth
W.  Thus the single $U$ limit could artificially push the system towards an
insulating state, or at least, increases the
localized characters of the $f$-electrons. To remedy this
situation, without increasing the level of complexity of the solution,
would be to replace the single $U$ limit by a single
$U-J$, since we have seen above that there is always an exchange term
in the density-density approximation for the $j m_j$ basis. In
that case, taking the overall average exchange $\overline{J}=0.323
eV$ of Eq.~(\ref{Jbar}), one would obtain a single peak located at
36.77 eV, which is in better agreement with more elaborated
results, and especially very close to the central maximum of the
density-density result.

%Here,
%\newpage

\section{\label{s:app} Atomic Temperature Green's function }

As mentioned in the introduction, one cannot neglect a
correct description of the atomic aspects of strongly
correlated electrons systems. In this part we concentrate on the
atomic (i.e., local) Green's function of an atom, which will be
later embedded in a solid. The results presented in this section can
be seen as a first step, with further hybridization to the
rest of the medium to be be added later as, for example,
is done in the DMFT
approach. We start from the definition of (imaginary time)
temperature Green function :
 \beq \EuScript{G}_k(\tau)=-\langle
T_{\tau} c_k(\tau) c_{k}^{\dagger} (0)\rangle\;,
 \label{eq2.1.1} \eeq
where $c_{k}^{\dagger}$ is a creation operator in a state $k$, which is,
in the present case, one of the 14 atomic orbitals of the $j m_j$
basis for $\ell=3$.  Based on the antiperiodic property \beq
\EuScript{G}_k(\tau +\beta)=-\EuScript{G}_{k}(\tau)\;,
\label{eq2.1.2}
\eeq
where ${1}/{\beta}$ is the temperature, one can write
the Fourier expansion of $\EuScript{G}_{k}(\tau)$ as
 \beq \EuScript{G}_{k}(\tau)=\frac{1}{\beta}\sum_{i \omega_n}
 e^{-i \omega_n \tau}\EuScript{G_{k}}(i \omega_n )\;,
 \label{eq2.1.3}
  \eeq
with the Fourier transform
\beq \EuScript{G}_{k}(i \omega_n )=\int_{0}^{\beta} e^{i \omega_n
\tau} \EuScript{G}_{k}(\tau)d\tau\;. \label{eq2.1.4} \eeq
The
Matsubara frequencies $\omega_n$ are given by:
\beq \omega_n=
(2n+1)\pi/\beta\;. \eeq

 If the eigenstates $ | a \rangle$ and their
eigenvalue $E_a$ are known from a diagonalization of the Hamiltonian,
then Eq.~(\ref{eq2.1.4}) has the following Lehmann
representation:
\beq \EuScript{G}_{k}(i \omega_n
)=\frac{1}{Z_G}\sum_{a,b}|\langle a |c_k| b
\rangle|^{2}\frac{e^{-\beta E'_a}+e^{-\beta E'_b}}{i \omega_n
+E'_a-E'_b }\;.
\label{eq2.1.5}
\eeq
Here $Z_G$ is the grand partition
function:
\beq
Z_G=Tr e^{-\beta (H-\mu N)}\;,
 \eeq
$E'_a=E_a- \mu N_a$,
and $\mu$, the chemical potential, is chosen to fix the average
number of particles with:
 \beq \sum_{k} \frac {1}{\beta} \sum_{i
\omega_n}
 e^{+i\omega_n 0^{+}} \EuScript{G}_{k}(i \omega_n )=\sum_{k} N_k \;.
 \label{eq2.1.6} \eeq
Eq.~(\ref{eq2.1.5}) can be used to define the spectral density; this
can be compared to photoemission experiments, and is given by
\beq \rho_k (\omega)= \frac{1}{Z_G}\sum_{a,b}|\langle a |c_k| b
\rangle|^{2} (e^{-\beta E'_a}+e^{-\beta E'_b})
\delta(\omega+E'_a-E'_b)\;. \label{eq2.1.7}
\eeq
Because the sum involves
discrete eigenstates for the atomic case, we add to $\omega$ a
small imaginary part so that the spectral density becomes a sum of
Lorentzians when the results are presented.
It should also be noted that
for solids the atomic Green's function tends to become a very good
approximation at sufficiently high temperature
(see, for example, the discussion of Fig.~2 in Ref.~\onlinecite{AMcMahan}).

\subsection{\label{GSingleUmodel}
Comparison of the single $U$ model with the density-density approximation}

Since all states of a given occupancy
$N$ are degenerate with eigenenergy $U {N(N-1)}/{2}$
when spin-orbit is neglected,
the single $U$ model has an analytical
expression for the Green function, which can be written:
%\widetext
%\begin{eqnarray}
\begin{align}
\EuScript{G}_{k}(i \omega_n )&=\frac{1}{Z_G}\sum_{N}
\frac{13!}{N!(13-N)!}
\nonumber \\
&\times
\frac{e^{-\beta
N[\frac{N-1}{2}U-\mu]}+e^{-\beta (N+1)[\frac{N}{2}U-\mu]}}{i
\omega_n +\mu-NU}\;.\label{GsinglU}
\end{align}
%\end{eqnarray}
%\endwidetext
This result, which can be presented in a form previously obtained
by McMahan
 et al.,~\cite{AMcMahan} shows evenly spaced values separated
 by $U$ in the poles of
 the Green function.

\begin{figure}[t!]
%\begin{center}
%\rotatebox{0}{\includegraphics[width=8cm]{fig3.eps}} \vspace{.5cm}
\centerline{\psfig{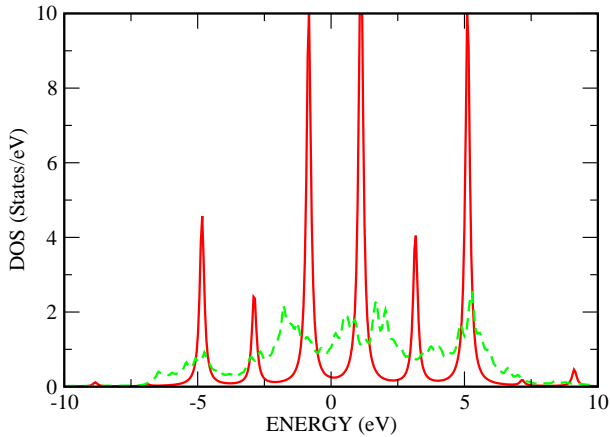}}
\caption{\label{jxz} (Color online) Spectral density of atomic Pu
at T=15800K, for single U limit (full line) and density-density
approximation (dashed line) respectively, including spin-orbit
coupling in both cases. }
%\end{center}
\end{figure}

When spin-orbit is switched on ($\chi \neq 0$), it is less
convenient to give an simple analytical formula, since the
degeneracy of the states is partially lifted. We present the
spectral density for this case in Fig. \ref{jxz}. One can still
see the effect of approximately evenly spaced values of the order
of $U$ for the poles, but additional structure also appears. Note
that the effect of temperature, here chosen to be 15800~K, allows
additional excited states to appear in the spectra. A lower
temperature would have completely quenched much of this additional
structure. In addition, this temperature has been used in some
recent DMFT Pu calculations to check the Quantum Monte Carlo (QMC)
DMFT codes against atomic-like Hubbard-I approximations
\cite{Zhu07}, since the atomic limit is a good approximation at
this temperature. The average $f$-occupancy is chosen to be about
5.6 in accordance with the result of Ref \cite{Zhu07}. The
comparison with the density-density approximation spectral density
(dashed line) in Fig.~\ref{jxz} clearly shows that the single $U$
approximation is an oversimplified model for the correlation
effects, and misses many of the features of even the
density-density results, and misplaces some of the peaks.

%\newpage

%\newpage
\subsection{\label{ss:cbnd} Comparison of the density-density approximation
with exact results}

\begin{figure}[h]
%\begin{center}
%\rotatebox{0}{\includegraphics[width=8cm]{fig4.eps}} \vspace{.5cm}
\centerline{\psfig{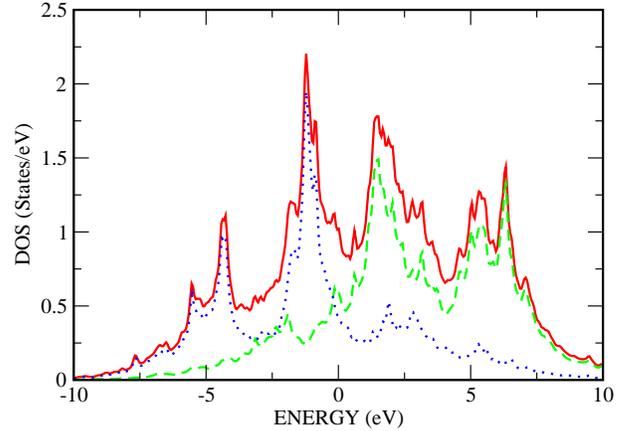}}
\caption{\label{rho(omega)sflip} (Color online) Exact results,
which include spin-flip terms in Hamiltonian, for the total (full
line), and 5/2 (dotted line) and 7/2 (dashed line) projected
spectral densities of atomic Pu at T=15800K, including spin-flip
terms in Hamiltonian}
%\end{center}
\end{figure}

In Fig.~\ref{rho(omega)sflip} we show the exact results for
the spectral density (within the basis we have chosen).
These include the spin-flip terms and involve exact diagonalization
of the local atomic Hamiltonian.  In this figure we also include the
decomposition of the spectra into its 5/2 and 7/2 contributions, with their
respective weights of 6 and 8.  As expected, the spin-orbit
splitting has displaced the gross features of the 5/2 density
downward with respect to those of the 7/2 features.
These spectral densities have much more
structure in them than the density-density result.

%\newpage
%\newpage

\begin{figure}[h]
%\begin{center}
%\rotatebox{270}
%{\includegraphics[width=8cm]{fig5.eps}} \vspace{.5cm}
\centerline{\psfig{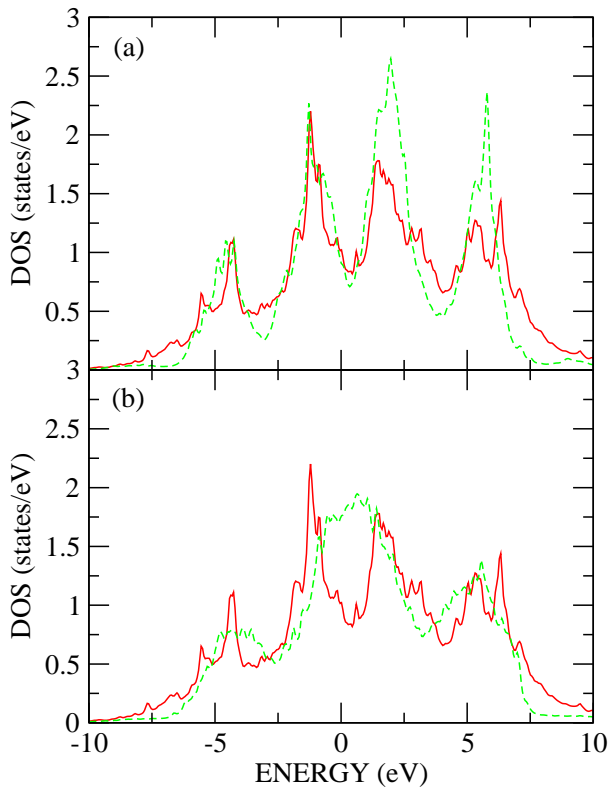}}
\caption{\label{sflip-dens-dens} (Color online) Comparison of the
spectral density of atomic Pu, including spin-flip terms in
Hamiltonian (full line), with the density-density approximation
result (dashed line). The upper panel (a) provides results for the
density-density approximation for j-j coupling ($n_j n_{j'}$)
whereas the bottom panel (b) is for the case of $ls$ coupling
($n_{ls} n_{l's'}$). Note that for the latter case the full matrix
must be diagonalized since the spin-orbit coupling term is not
diagonal in the $ls$ basis. The spin-flip case is reproduced in
both panels for comparison.}
%\end{center}
\end{figure}

In Fig.~\ref{sflip-dens-dens} we compare the spectral
density for the density-density approximation with
the exact results for both $j$-$j$ and $ls$ coupling.
Both show a similar overall gross structure of the spectrum,
which is located in approximately the same energy range.
However, the $j$-$j$ coupling clearly does a much superior
job in reproducing the actual peak structure in the spectrum.
In addition, the excellent agreement between the
$j$-$j$ density-density approximation and exact spectral
densities suggests that in many situations the $j$-$j$
density-density approximation may be an adequate
approximation to more refined treatments of the
Coulomb effects for Pu.
It is worth mentioning that, when
all terms are retained as here, i.e., spin-orbit and full
electrostatic interaction, the results do not depend on the
basis chosen for the representation; the local Hamiltonian
that would have been solved from the $LS$ basis would have
had the same final spectral density. Thus the similarity we
stress between the full result and the density-density
approximation is \textit{not} an artifact of the choice of
representation, namely the $jm_j$ basis.

\section{\label{s:ccl} Summary and conclusion}
We have presented a study of the atomic electronic
structure of the $f$-electron elements in the presence of spin-orbit coupling,
and applied this to plutonium. We have emphasized the role of
electronic interactions in a $jm_j$ basis. For this purpose, we have
derived a general analytical expression for the matrix elements of the
interaction in this basis; some of these that can be used in a
density-density approximation have been tabulated, since they could
be useful for possible calculations in solids. We have used
these to study the effect of Coulomb interactions on the eigenvalue
and spectral functions. Diagonalizations of
this Hamiltonian for different occupancies and computations of
spectral density, have been performed in various approximations, including
single $U$, single $U-J$, density-density ,and, finally, the full
interaction matrix retaining all spin-flip terms.
We have found that the density-density approximation,
in the $jm_j$ basis,  which neglects off-diagonal terms (i.e.,
spin-flip terms) in the interaction matrix, gives excellent
agreement with the full interaction.

The high accuracy of the $j$-$j$ density-density approximation
should make it very useful in electronic-structure calculations
for solids.  For example, it is a very tractable method
for the Hirsch-Fye
algorithm of the QMC solver in DMFT, and should makes the
application of the Gutzwiller method much
easier.

\section{Acknowledgments}
This work was carried out under the auspices of the National
Nuclear Security Administration of the U.S. Department of Energy
at Los Alamos National Laboratory under Contract No.
DE-AC52-06NA25396. Financial supports from LANL Theoretical
Division (CNLS and T-11 group) as well as DGA (Delegation Generale
pour l'Armement Contract No. 07.60.028.00.470.75.01) are
gratefully acknowledged. J.-P. J also acknowledges the LANL group
for his warm hospitality during his stay in Los Alamos, USA.

%\newpage
\appendix*
\section{Expression of the general matrix element in the $jm_j$ basis}
We demonstrate now formula (\ref{Vspflip}). The general expression
element of the Coulomb interaction in $j m_j$ basis is
$V_{1234}=\langle j_1 M_1  j_2 M_2| V| j_3 M_3 j_4
M_4\rangle$. In this appendix to avoid too heavy notations, we
will use the notation $| j M\rangle$ for an element of the
$jm_j$ basis and $| \ell m \sigma\rangle$ to designate the
state $| \ell \, m_{\ell} \, s \, m_s \rangle$ of the $\ell m_{\ell}$
basis. $j_1$ to $j_4$ could be one of either $\ell+\frac{1}{2}$ or
$\ell-\frac{1}{2}$ total momentum quantum number ($\ell$ being
well defined here). By inserting four times the Clebsch-Gordan
(CG) expansion:
\beq | j M\rangle=\sum_{m \sigma}| \ell  m \sigma\rangle
\langle\ell  m \sigma| j M\rangle \label{A1}\eeq
in the definition of $V_{1234}$, where $\langle\ell
m \sigma|j M\rangle $ is a shorthand for the (chosen real) CG
coefficient $\langle\ell m \frac{1}{2} \sigma|(\ell
\frac{1}{2}) j M\rangle$, one obtains:
%\widetext
\begin{eqnarray}
V_{1234}&=& \sum_{m_1 \sigma_1 m_2 \sigma_2 m_3 \sigma_3 m_4
\sigma_4} \langle\ell m_{1} \sigma_{1}|j_{1} M_{1}\rangle
\langle\ell m_{2} \sigma_{2}|j_{2}  M_{2}\rangle \nonumber\\
&&\times \langle\ell m_{3} \sigma_{3}|j_{3} M_{3}\rangle
\langle\ell  m_{4} \sigma_{4}|j_{4} M_{4}\rangle \nonumber\\
&&\times \langle\ell m_{1} \sigma_{1} \, \ell m_{2} \sigma_{2}|V|\ell
m_{3} \sigma_{3} \, \ell m_{4} \sigma_{4}\rangle \;. \nonumber\\
\label{A2}
\end{eqnarray}
%\endwidetext
The matrix element of interaction in the $\ell m_{\ell}$ basis
appearing in this last formula can be expressed as:
%\widetext
%\begin{eqnarray}
\begin{align}
\langle\ell m_{1}\sigma_{1}&\ell m_{2}\sigma_{2}|V|\ell
m_{3} \sigma_{3} \ell m_{4} \sigma_{4}\rangle =
\delta_{\sigma_{1}\sigma_{3}}\delta_{\sigma_{2}\sigma_{4}}
\nonumber \\
&\times \int  dr d\Omega dr' d\Omega' Y_{\ell m_{1}}^{*}(\Omega)Y_{\ell
m_{2}}^{*}(\Omega') \nonumber \\
&\times |R_{n\ell}(r)|^2
\frac{1}{|r-r'|}|R_{n\ell}(r')|^2 Y_{\ell m_{3}}(\Omega)Y_{\ell
m_{4}}(\Omega')\;. \nonumber \\
\label{A3}
\end{align}
%\end{eqnarray}
%\endwidetext

>From Eq.~(\ref{coulexpans}), the above integral can be
conveniently cast into a sum of a product of one radial integral
(identified as the Slater integral (\ref{SlaterInt})) and two
angular integrals:
%\widetext
%\begin{eqnarray}
\begin{align}
\sum_{k=0}^{\infty}\sum_{q=-k}^{k}&\frac{4\pi}{2k+1}\int  dr dr'
|R_{n\ell}(r)|^2
\frac{r_{<}^k}{r_{>}^{k+1}}|R_{n\ell}(r')|^2 \nonumber \\
&\times \int d\Omega Y_{\ell m_{1}}^{*}(\Omega)Y_{k
q}^{*}(\Omega)Y_{\ell m_{3}}(\Omega) \nonumber \\
&\times \int d\Omega'Y_{\ell m_{2}}^{*}(\Omega')Y_{k q}(\Omega')Y_{\ell
m_{4}}(\Omega')\;. \label{A4}
\end{align}
%\end{eqnarray}
%\endwidetext

With conjugation relation $Y_{\ell m}^{*}(\Omega)=(-)^{m}Y_{\ell
-m}(\Omega)$, each one of the two angular integrals can calculated
from a well-known identity giving the integral of the product of
three spherical harmonics:
%\widetext
%\begin{eqnarray}
\begin{align}
\int d\Omega Y_{\ell_1 m_1 }&(\Omega)Y_{\ell_2 m_2
}(\Omega)Y_{\ell_3 m_3
}(\Omega)=
(-)^{m_3} \nonumber \\
&\times \sqrt{\frac{(2\ell_1+1)(2\ell_2+1)}{4 \pi(2\ell_3+1)}} \nonumber \\
&\times \langle \ell_{1} 0 \ell_2 0| \ell_3 0\rangle
\langle \ell_1 m_1 \ell_2 m_2| \ell_3 \, {-}m_3\rangle\;. \label{A5}
\end{align}
%\end{eqnarray}
%\endwidetext

Application of this last identity with the selection rule for CG
coefficients $\langle \ell_1 m_1 \ell_2 m_2| \ell_3 m_3\rangle$
which vanish unless $m_3=m_1+m_2$, enables to rewrite (\ref{A3}):
%\widetext
%\begin{eqnarray}
\begin{align}
\langle\ell m_{1} &\sigma_{1} \, \ell  m_{2}\sigma_{2}|V|\ell
m_{3} \sigma_{3} \, \ell m_{4} \sigma_{4}\rangle =
(-)^{m_{1}-m_{2}} \nonumber \\
&\times \delta_{\sigma_{1}\sigma_{3}}
\delta_{\sigma_{2}\sigma_{4}}\delta_{m_{1}+m_{2},m_{3}+m_{4}}
\sum_{k=0}^{2\ell}
F^{(k)}\frac{(2\ell+1)^2}{(2k+1)^2} \nonumber \\
&\times \langle \ell \, 0 \, \ell \, 0| k \, 0\rangle^{2}
\langle \ell \, {-}m_{1} \,  \ell \, m_{3}| k \, m_{3}{-}m_{1}\rangle \nonumber \\
&\times \langle \ell \, {-}m_{2} \, \ell \, m_{4}| k \, m_{4}{-}m_{2}\rangle \;.
\label{A6}
\end{align}
%\end{eqnarray}
%\endwidetext
The fact that CG $\langle \ell\; 0 \; \ell \;0| k\; 0\rangle$
vanishes unless $2\ell+k$ is even and $k$ preserves the well-known
triangle condition, namely $0 \leq k \leq 2\ell$, enables to
restrain summation to even values of $k$ from 0 to $2\ell$,
\textit{i.e.} limited to $k=6$ for $f$ states, requiring only
Slater integrals $F^{(0)}$ to $F^{(6)}$, given in Table I for
Plutonium. Replacement of result (\ref{A6}) into expression
(\ref{A2}), associated with selection rule for CG coefficients
$\langle\ell\; m\;\sigma|j\; M\rangle $ which vanish unless
$M=m+\sigma$, and final use of identity for $\pm\frac{1}{2}$
spins, $ (-)^{\sigma_{1}-\sigma_{2}}=(2\sigma_{1})(2\sigma_{2})$,
one derives directly expression:
%\widetext
%\begin{eqnarray}
\begin{align}
V_{1234}&=\delta_{M_{1}+M_{2},M_{3}+M_{4}}(-)^{(M_{3}-M_{2})}
\nonumber \\
&\times \sum_{k=0}^{2\ell}
F^{(k)}\frac{(2\ell+1)^2}{(2k+1)^2}
\langle \ell \, 0 \, \ell \, 0| k \, 0\rangle^{2}
\nonumber \\
&\times
\sum_{\sigma_1 \sigma_2}(2\sigma_{1})(2\sigma_{2})
\langle\ell \, M_{1} {-} \sigma_{1} \, \sigma_{1}|j_{1} \, M_{1}\rangle
\nonumber \\
&\times \langle \ell \, M_{2} {-} \sigma_{2} \, \sigma_{2}|j_{2} \, M_{2}\rangle
%\nonumber \\
%&\times
\langle \ell \, M_{3} {-} \sigma_{1} \, \sigma_{1}|j_{3} \, M_{3}\rangle
\nonumber \\
&\times
\langle \ell \, M_{4} {-} \sigma_{2} \, \sigma_{2}|j_{4} \, M_{4}\rangle
\nonumber \\
&\times
\langle \ell \,  {-} (M_{1} {-} \sigma_{1}) \, \ell, M_{3} {-} \sigma_{1}| k \, M_{3} {-} M_{1}\rangle
\nonumber \\
&\times
\langle \ell \,  {-} (M_{2} {-} \sigma_{2}) \, \ell \, M_{4} {-} \sigma_{2} |
k \, M_{4} {-} M_{2}\rangle\;,
\label{A7}
\end{align}
%\end{eqnarray}
%\endwidetext
\noindent from which one obtains expression (\ref{Vspflip}) when we make an
identification of definition~(\ref{B}) for auxiliary function
$B_{k}^{\ell}(jm;j'm')$.

Application of formula~(\ref{Vspflip})  to direct and exchange
cases lead us to arrive at  Eqs.~(\ref{aso}) and (\ref{bso}),
respectively. From these expressions and the CG related symmetries
properties of function $B_{k}^{\ell}(jm;j'm')$ given
in~(\ref{B}), one can give the symmetry relations that $a$ and $b$'s
obey, reducing the number entries of Tables II and III. The matrix
elements not given in those Tables can be consequently obtained
using the following symmetries:
\begin{eqnarray}
 a_{k}^{\ell}(j, |m|; j', |m'|)&=&a_{k}^{\ell}(j m; j' m')\;,
\label{aso2}
\end{eqnarray}
 and
\begin{eqnarray}
b_{k}^{\ell}(j m; j' m')&=&b_{k}^{\ell}(j \,  {-}m;j' \,  {-}m')\;,\nonumber\\
b_{k}^{\ell}(j \,  {-}m; j', m')&=&b_{k}^{\ell}(j,m; j' \,  {-}m')\;.
\label{bso2}\end{eqnarray}

%\newpage

%\section{Values for direct matrix element of interaction in $jm_j$ basis}

%Table II
%\setlongtables
\begin{longtable}{crrrrrc}
%\begingroup
%\squeezetable
%\begin{table}[H]
\caption{\label{tab:direct}
Values of $a_{k}^{\ell}(jm; j'm')$ for the direct interaction matrix
element $\langle jm j' m'|V|j m j' m'\rangle=\sum_{k}a_{k}^{\ell}
(j m; j'm')F^{(k)}$ for $\ell=3$ in the $j m_j$ basis.}
%\begin{ruledtabular}
%\begin{tabular}{crrrrrc}
\endfirsthead
\hline\hline
$j$  &$|m_{j}|$ &  $j'$   &$|m_{j}'|$&$F^{(2)}$  &$F^{(4)}$&$F^{(6)}$     \\
 \hline
7/2     & 7/2      &  7/2    &  7/2    &   49/441   & 49/5929 &  25/184041   \\
          & 7/2      &         &  5/2    &    7       &  -91    &   -125       \\
          & 7/2      &         &  3/2    &  -21       &  -21    &    225       \\
%\vspace{0.2cm}
          & 7/2      &         &  1/2    &  -35       &   63    &   -125       \\
          & 5/2      &         &  5/2    &    1       &   169   &    625       \\
          & 5/2      &         &  3/2    &   -3       &    39   &  -1125       \\
%\vspace{0.2cm}
          & 5/2      &         &  1/2    &   -5       &  -117   &    625       \\
          & 3/2      &         &  3/2    &    9       &   9     &   2025       \\
%\vspace{0.2cm}
          & 3/2      &         &  1/2    &   15       & -27     &  -1125       \\
          & 1/2      &         &  1/2    &   25       &  81     &    625       \\
\hline
%----------------------------------------------------------------------------------------
7/2       & 7/2      &  5/2    &  5/2    &   70/735   & 7/1617  &    0         \\
          & 7/2      &         &  3/2    &  -14       & -21     &    0         \\
%\vspace{0.2cm}
          & 7/2      &         &  1/2    &  -56       &  14     &    0         \\
          & 5/2      &         &  5/2    &   10       & -13     &    0         \\
          & 5/2      &         &  3/2    &   -2       &  39     &    0         \\
%\vspace{0.2cm}
          & 5/2      &         &  1/2    &   -8       & -26     &    0         \\
          & 3/2      &         &  5/2    &  -30       &  -3     &    0         \\
          & 3/2      &         &  3/2    &    6       &   9     &    0         \\
%\vspace{0.2cm}
          & 3/2      &         &  1/2    &   24       &  -6     &    0         \\
          & 1/2      &         &  5/2    &  -50       &    9    &    0         \\
          & 1/2      &         &  3/2    &   10       &  -27    &    0         \\
          & 1/2      &         &  1/2    &   40       &   54    &    0         \\
\hline
%----------------------------------------------------------------------------------------
5/2       & 5/2      &  5/2    &  5/2    &  100/ 1225 & 1/441   &    0         \\
          & 5/2      &         &  3/2    &   -20      &  -3     &    0         \\
%\vspace{0.2cm}
          & 5/2      &         &  1/2    &   -80      &   2     &    0         \\
          & 3/2      &         &  3/2    &    4       &   9     &    0          \\
%\vspace{0.2cm}
          & 3/2      &         &  1/2    &   16       &  -6     &    0          \\
          & 1/2      &         &  1/2    &   64       &   4     &     0        \\
%\end{tabular}
%\end{ruledtabular}
%\end{table}
\hline\hline
\end{longtable}
%\endgroup

%\newpage
%\section{Values for exchange matrix element of interaction in $jm_j$ basis}

%\begingroup
%\squeezetable
%\begin{table}[H]
\begin{longtable}{crrrrrc}
\caption{\label{tab:exchange}
Values of $b_{k}^{\ell}(j m; j'
m')$ for the exchange interaction matrix element $\langle j m
j' m'|V|j'm' j m\rangle=\sum_{k}b_{k}^{\ell}(j m; j'
m')F^{(k)}$ for $\ell=3$ in the $j m_j$ basis.}
%\begin{ruledtabular}
%Table III  \\
%\begin{tabular}{crrrrrc}
\endfirsthead
\hline\hline
     $j$  &$m_{j}$   &  $j'$   &$m_{j}'$ &$F^{(2)}$  &$F^{(4)}$&$F^{(6)}$     \\
     \hline
7/2       &$ \pm7/2$   &  7/2    & $\pm7/2$  &  49/441   & 49/5929 & 25/184041    \\
          & $\pm7/2$   &         &$ \pm5/2$  &    42     & 140     &  150         \\
          & $\pm7/2$   &         & $\pm3/2 $ &    14     & 210     &  500         \\
          & $\pm7/2$   &         &$ \pm1/2$  &     0     & 196     & 1200         \\
          & $\pm7/2$   &         &$ \mp1/2$  &     0     & 98      & 2250         \\
          & $\pm7/2$   &         &$ \mp3/2$  &     0     &    0    & 3300         \\
          & $\pm7/2$   &         &$ \mp5/2 $ &     0     &    0    & 3300         \\
%\vspace{0.2cm}
          & $\pm7/2$   &         &$ \mp7/2 $ &     0     &    0    &    0         \\
          & $\pm5/2 $  &         &$ \pm5/2$  &     1     & 169     &  625         \\
          & $\pm5/2$   &         & $\pm3/2$  &    32     &  60     & 1400         \\
          &$ \pm5/2$   &         & $\pm1/2$  &    30     &  2      & 2100         \\
          & $\pm5/2 $  &         &$ \mp1/2 $ &     0     & 112     & 2100         \\
          & $\pm5/2$   &         &$ \mp3/2$  &     0     & 210     & 1050         \\
%\vspace{0.2cm}
          &$ \pm5/2$   &         & $\mp5/2$  &     0     &    0    &    0         \\
          & $\pm3/2  $ &         & $\pm3/2$  &     9     &    9    & 2025         \\
          &$ \pm3/2  $ &         & $\pm1/2 $ &    60     &  108    & 1750         \\
          & $\pm3/2$   &         & $\mp1/2$  &    40     &   96    &  700         \\
%\vspace{0.2cm}
          & $\pm3/2$   &         & $\mp3/2 $ &     0     &    0    &    0         \\
          &$ \pm1/2  $ &         &$ \pm1/2 $ &    25     &   81    &   625        \\
          &$ \pm1/2$   &         & $\mp1/2$  &     0     &    0    &    0         \\
          \hline
%\end{tabular}
%\end{center}
%________________________________________________________________________________________
%\begin{center}
%Table III (\textit{continued}) \\
%\caption{\label{Table III}Table III (\textit{continued)}
%\vspace{0.5cm}
%\begin{tabular}{|crrrrrc|}   \hline
%     $j$  &$m_{j}$   &  $j'$   &$m_{j}'$ &$F^{(2)}$  &$F^{(4)}$&$F^{(6)}$     \\ \hline
%
7/2       & $\pm7/2$   &   5/2   & $\pm5/2$  & 175/11025 &210/53361& 25/184041    \\
          & $\pm7/2 $  &         & $\pm3/2$  &   140     &  756    &  200         \\
          &$ \pm7/2$   &         &$ \pm1/2$  &     0     & 1323    &  900         \\
          & $\pm7/2$   &         & $\mp1/2$  &     0     & 1176    & 3000         \\
          &$ \pm7/2 $  &         &$ \mp3/2 $ &     0     &    0    & 8250         \\
%\vspace{0.2cm}
          & $\pm7/2$   &         & $\mp5/2$  &     0     &    0    &19800          \\
          &$ \pm5/2$   &         & $\pm5/2$  &   150     &  600    &  150          \\
          & $\pm5/2$   &         & $\pm3/2 $ &     5     & 1014    &  875          \\
          & $\pm5/2$   &         &$ \pm1/2 $ &   160     &  486    & 2800          \\
          & $\pm5/2$   &         &$ \mp1/2 $ &     0     &   21    & 6300          \\
          &$ \pm5/2$   &         &$ \mp3/2  $&     0     & 1344    &10500        \\
%\vspace{0.2cm}
          & $\pm5/2$   &         &$ \mp5/2 $ &     0     &     0   &11550          \\

          &$ \pm3/2$   &         &$ \pm5/2$  &    75     & 1000    &  525          \\
          & $\pm3/2$   &         &$ \pm3/2$  &    90     &  640    & 2250          \\
          &$ \pm3/2$   &         & $\pm1/2 $ &    30     &    1    & 5250          \\
          &$ \pm3/2$   &         &$ \mp1/2$  &   120     &  578    & 8400          \\
          & $\pm3/2$   &         &$ \mp3/2 $ &     0     &  686    & 9450          \\
%\vspace{0.2cm}
          & $\pm5/2 $  &         & $\mp5/2$  &     0     &  560    & 6300          \\

          &$ \pm1/2$   &         & $\pm5/2$  &    20     & 1200    & 1400          \\
          & $\pm1/2$   &         & $\pm3/2 $ &   121     &  120    & 4375          \\
          & $\pm1/2$   &         & $\pm1/2$  &    12     &  300    & 7500          \\
          &$ \pm1/2  $ &         & $\mp1/2  $&    98     &  735    & 8750          \\
          &$ \pm1/2$   &         & $\mp3/2$  &    64     &   60    & 7000          \\
          & $\pm1/2$   &         & $\mp5/2 $ &     0     & 1050    & 3150          \\
          \hline
%________________________________________________________________________________
5/2     & $\pm5/2$   &  5/2    & $\pm5/2$  &  100/1225 &  1/441  &   0    \\
          &$ \pm5/2$   &         & $\pm3/2$  &   120     &    4    &   0           \\
          &$ \pm5/2$   &         & $\pm1/2$  &    60     &    9    &   0           \\
          & $\pm5/2$   &         & $\mp1/2$  &     0     &   14    &   0           \\
          &$ \pm5/2$   &         &$ \mp3/2$  &     0     &   14    &   0           \\
%\vspace{0.2cm}
          & $\pm5/2$   &         &$ \mp5/2$  &     0     &    0    &   0           \\

          &$ \pm3/2$   &         &$ \pm3/2 $ &     4     &    9    &   0           \\
          &$ \pm3/2$   &         &$ \pm1/2 $ &    48     &   10    &   0           \\
          & $\pm3/2 $  &         &$ \mp1/2$  &   108     &    5    &   0           \\
%\vspace{0.2cm}
          & $\pm3/2$   &         &$ \mp3/2$  &     0     &    0    &   0           \\

          &$ \pm1/2$   &         & $\pm1/2 $ &    64     &    4    &   0           \\
          &$ \pm1/2$   &         & $\mp1/2 $ &     0     &    0    &   0           \\

%\end{tabular}
%\end{ruledtabular}
%\end{table}
%\endgroup
\hline\hline
\end{longtable}

\end{document}